\documentstyle[12pt]{article}
\begin{document}
\setcounter{page}{1}
\renewcommand{\thesection}{\Roman{section}}
\begin{center}
{\Large\bf The rest mass}\\
\vspace{1cm}
\normalsize\it
Arup Roy\dag\\
Department of Physics\\
Scottish Church College\\
1 \& 3 Urquhart Square\\
Kolkata - 700006\\
India\\
\end{center}
\hspace*{1.5cm}
A relation connecting the {\it{rest mass}} and
{\it{separation of events in space-time continuum}} is suggested
and the idea of Compton scattering is used as a method for the
determination of rest mass. An experiment involving collision of
photons resulting in creation of rest mass is discussed theoretically
in order to illustrate the connection formula.\\[1cm]
\dag{\it{e.mail: ary@vsnl.net}}\\
\newpage
\hspace*{2cm}
Historically the rest mass first appears in the special theory of
relativity as a mathematical by-product when proper momentum was
defined and the conservation of 4-momentum was taken into
consideration~[1]. P. G. Bergmann~[1] in course of discussion on
the relation between energy and mass in relativistic mechanics observed
that if kinetic energy of a mechanical system decreases, at least some
of the rest masses of the constituent system must increase. This
so-called mass-energy equivalence accounts for mass defect in atomic
nuclei, disintegration energy of decay processes and so on and so
forth. Rest masses change appreciably when interaction energies have
the order of magnitude of the rest energies. Till to - date only a few
attempts are made to determine the rest mass experimentally; Jian Qi
Shen~[2] performed experiments on photon rest mass. Though an attempt
has been made by Donald Chang~[3] to investigate the wave properties
corresponding to rest mass, no real effort has yet been made to resolve
the fundamental question regarding the status of the {\it{rest mass}}
and to consider whether any basic principle is involved that inherently
restricts its determination.\\
\hspace*{2cm}
It is quite obvious that following definition~[4] of energy in special
theory of relativity and principle of energy and momentum framed therein
the loss of rest mass accounts for the production of energy and vice
versa. However the transfer of energy ($\triangle{E}$) is connected to
time interval ($\triangle{t}$) needed for the transfer to take place by\\
\begin{equation}
\triangle{E}\triangle{t} \geq \hbar
\end{equation}
and ($\triangle{E}$) can be measured with the available measuring
apparatus. But how can one measure {\it{rest mass}} ? Obviously it
cannot be done directly because any such effort will disturb the system
to the extent of changing the rest energy $E$ to some value $E$ $+$
$\triangle{E}$. One possible way is to consider Compton effect where
the theory assumes the particle (scatterer) to be free and at rest
before collision with the photon takes place. The Compton shift is
given by\\
\begin{equation}
\lambda - \lambda_{0} = (h/{mc})(1 - cos\theta)
\end{equation}
where $\lambda$ and $\lambda_{0}$ are the wavelengths respectively of
scattered and incident waves, $m$ is the rest mass of the particle and
$\theta$ is the angle of scattering. At a particular value of $\theta$,
the measurement of the Compton shift can give an estimate of the
{\it{rest mass}}. If one employs the expression (2) to find $m$, then
it becomes clear that the limit to measure $m$ accurately depends on
the limit to which $\lambda$ and $\lambda_{0}$ can be resolved. One can
use Rayleigh's criterion of resolution to get the maximum value of
rest mass that can be measured accurately, at least in principle.\\
\hspace*{2cm}
Now, the resolving power of an optical system is $R$ =
$\lambda_{0}/(\lambda - \lambda_{0})$, following symbols of
expression~(2). The determination of rest mass depends on clear
identification of the peaks for the two waves (incident and scattered).
For this to happen the shift in wavelength must be at least
$\lambda_{0}/R$, which leads to the inequation\\
\begin{equation}
m \leq hR(1 - cos\theta)/(c\lambda_{0})
\end{equation}
In other words, $m$ must have an upper limit.\\
\hspace*{2cm}
Now, let us consider the following process (ref. Fig. 1) {\it{viz.}}
collision of two oppositely directed photons resulting in the creation
of rest mass $m$, which in terms of usual symbols of the equation for
the conservation of $4$-momentum can be written as\\
\hspace*{3cm}
$(h\nu/c)(1, \vec{k})$ $+$ $(h\nu/c)(1, -\vec{k})$ = $(2mc, \vec{0})$,\\
$\vec{k}$ being the unit vector along the momentum of `photon 1'.
The Fig. 1 describes the process in the space-time continuum (only
one space dimension is shown). Thus,\\
\hspace*{3cm} $2h\nu/c$ = $2mc$ \hspace*{0.25cm} {\it{i.e.}}
$h\nu$ = $mc^{2}$;
\hspace*{0.25cm}and $(h\nu/c) \vec{k}$ $-$ $(h\nu/c) \vec{k}$ = $\vec{0}$\\
In this particular example, therefore, change $\triangle{m}$ in
rest mass is of the order of $h\nu/c^{2}$ which is equal to
$h/(c\lambda_{m})$ where $\lambda_{m}$ is the wavelength of either
of the photons.
 Let
$\triangle{s}$ denote an infinitesimally small interval separating
events {\it{viz.}}
collision of photons and creation of mass at rest, $s$ being given by\\
\hspace*{3cm} $(\triangle{s})^{2}$ = $c^{2}(\triangle\tau)^{2}$
$-$
$|\triangle\vec{r}|^{2}$\\
(the terms have their usual significance).\\
\newpage
\setlength{\unitlength}{1mm}
\begin{picture}(50,70)
\put(2,-49){\vector(2,0){150}} \put(152,-52){\makebox(0,0){$x$}}
\put(11,-52){\makebox(0,0){$-a$}}
\put(20,-46){\makebox(0,0){$45^0$ }} \put(12,-49){\line(1,1){64}}
\put(76,-49){\dashbox{0.5}(0.2,63)} \put(76,15){\line(0,3){20}}
\put(80,25){\makebox(0,0){$ct$}} \put(76,-52){\makebox(0,0){0}}
\put(140,-49){\line(-1,1){64}} \put(140,-52){\makebox(0,0){$a$}}
\put(132,-46){\makebox(0,0){$45^0$}} \put(4,40){\vector(3,0){30}}
\put(20,35){\makebox(0,0){photon 1}}
\put(20,45){\makebox(0,0){$h\nu$}} \put(138,40){\vector(-3,0){30}}
\put(125,35){\makebox(0,0){photon 2}}
\put(125,45){\makebox(0,0){$h\nu$}}
\end{picture}\\[6cm]
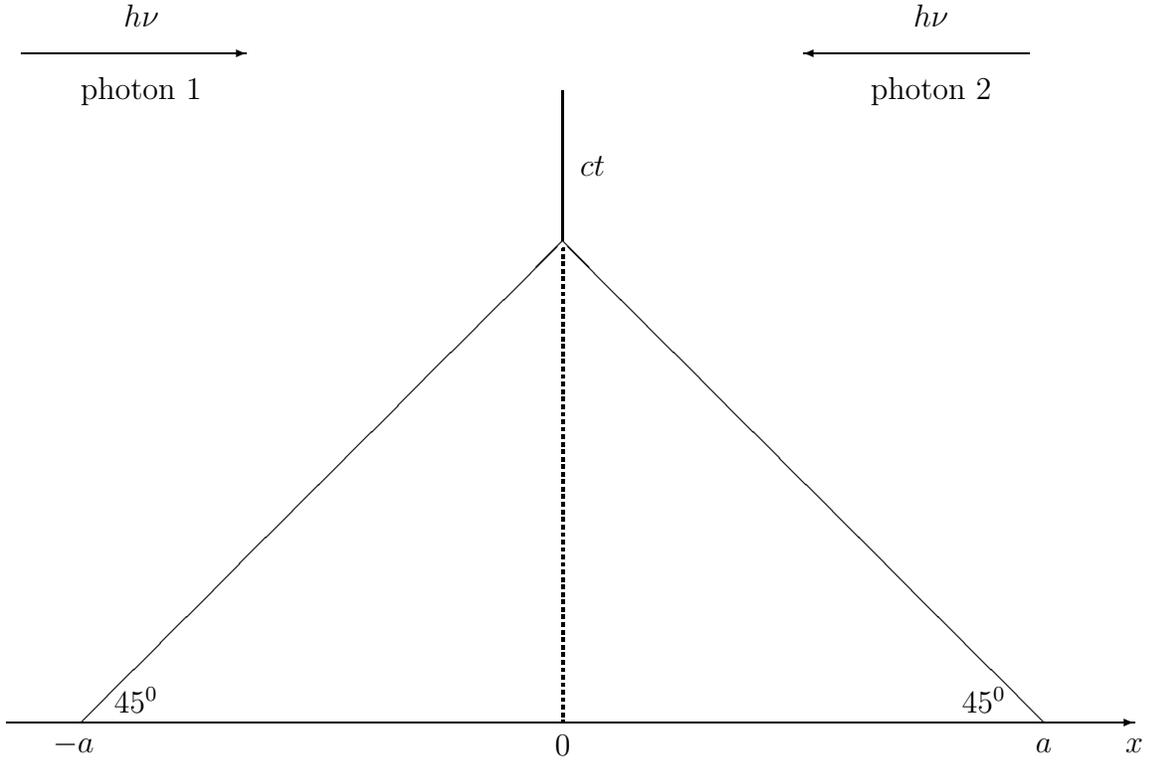
Fig. 1 Space-time diagram to illustrate the process where two
photons projected simultaneously from $x$ = $a$ and $x$ = $-a$,
meet at $x$ = 0 to form a point mass at rest.
\\
\\

The separation $\triangle{s}$ in this example should be of the
order of $\lambda_{m}$. Thus,
\begin{equation}
\triangle{m}\triangle{s} \sim \hbar/c
\end{equation}
In order to include the idea expressed in (3), expression (4) should be
rewritten as
\begin{equation}
\triangle{m}\triangle{s} \leq \hbar/c
\end{equation}
The very description that mass is at rest, makes one conceive of
localization (in space-time) of the system, and this is what is
precisely given in (5). For the sake of an application if we substitute
in (5) the Planck mass ($10^{19}$ $GeV$) for $\triangle{m}$, we get
for $\triangle{s}$ a value which is of the order of $10^{-33}$ $cm$,
the Planck length.\\
\hspace*{2cm}
For time like separation of events it might well be that with respect to
an observer $|\triangle\vec{r}|$ = $0$; then $\triangle{s}$ =
$c\triangle\tau$ and in the context of creation of mass (as discussed
above in connection with the problem of two photons) the world line of
the physical system will obviously be traced in such a way that
$\triangle\tau$ becomes positive, otherwise `causality' will be
violated. This $\triangle\tau$ signifying the separation in time of the
two events, must have a magnitude of the order of uncertainty
$\triangle{t}$ of relation (1) when $\triangle{E}$ becomes the energy
equivalent of $\triangle{m}$. Now for space-like separation it can be
that with respect to an observer $\triangle{\tau}$ = $0$, which means
that $(\triangle{s})^{2}$ $<$ $0$ {\it{i.e.}} $\triangle{s}$ is
imaginary. Thus it is evident from (5) that for space like separation
of events mass creation is not possible.\\
\hspace*{2cm}
Since loss of rest mass makes corresponding $\triangle{E}$ positive
and the gain in rest mass is always associated with loss in energy,
$\triangle{m}$ and $\triangle{E}$ must have opposite signs; the
uncertainty relation expressed by~(1) then follows simply from
relation~(5) for a time like separation of events when mass is being
created.\\


\end{document}